\documentclass[twocolumn,prb,aps,amsmath,amssymb,floatfix,superscriptaddress]{revtex4}
\usepackage{graphicx}
\usepackage{dcolumn}
\usepackage{bm}

\newcommand{\MYT}{Department of Applied Physics, Chalmers University of
Technology, SE--412 96 G\"oteborg, Sweden}
\newcommand{\BN}{Microtechnology and Nanoscience, MC2, Chalmers University of
Technology, SE--412 96 G\"oteborg, Sweden}

\begin{document}
\title{Potassium intercalation in graphite: A van der Waals 
density-functional study}

\author{Eleni Ziambaras}\affiliation{\MYT} 
\author{Jesper Kleis}\affiliation{\MYT}
\author{Elsebeth Schr\"oder}\affiliation{\MYT}
\author{Per Hyldgaard}\email{hyldgaar@chalmers.se}\affiliation{\MYT}\affiliation{\BN}

\date{April 1, 2007}

\begin{abstract}
Potassium intercalation in graphite is investigated by
first-principles theory. The bonding in the potassium-graphite 
compound is reasonably well accounted for by 
traditional semilocal density functional theory (DFT) calculations.
However, to investigate the intercalate formation energy from pure potassium atoms and
graphite
requires use of a description of the graphite interlayer binding and thus
a consistent account of the nonlocal 
dispersive interactions. This is included seamlessly with ordinary DFT
by a van der Waals density functional (vdW-DF) approach [Phys.\ Rev.\ Lett.\ \textbf{92},
246401 (2004)]. The use of the vdW-DF is found to
stabilize the graphite crystal, with crystal parameters in fair
agreement with experiments. 
For graphite and potassium-intercalated graphite 
structural parameters such as binding separation, layer binding energy, formation
energy, and bulk modulus 
are reported. Also the adsorption and sub-surface potassium absorption energies 
are reported. The vdW-DF description, compared with the traditional 
semilocal approach, is found to weakly soften the elastic response.
\end{abstract}

\maketitle

\section{Introduction}

Graphite with its layered structure is easily intercalated 
by alkali metals (AM) already at room temperature. 
The intercalated compound has two-dimensional layers of
AM between graphite layers,\cite{aronson,nixon1968,clarke2,k_intercalation,Di} 
giving rise to interesting
properties, such as superconductivity.\cite{supercond,jishi}
The formation of an AM-graphite intercalate
proceeds with adsorption of AM atoms on graphite and absorption
of AM atoms below the top graphite layer, after which
further exposure to AM atoms leads the AM intercalate compound.

Recent experiments\cite{exp1_interc,exp2_interc} on the
structure and electronic
properties of AM/graphite systems
use samples of graphite that are
prepared by heating SiC crystals to temperatures around
$\sim 1400^\circ$ C.\cite{forb} This heat-induced graphitization
is of great value for spectroscopic studies of
graphitic systems, since the resulting graphite overlayers are of excellent
quality.~\cite{sic_grafit} The nature of the bonding between the SiC surfaces 
and graphite has been explored experimentally with photoemission 
spectroscopy~\cite{forb2}
and theoretically~\cite{per_eleni_sic} with a van der Waals
density functional (vdW-DF) theory approach that accounts for the van der 
Waals (vdW) forces.~\cite{henrik,ijqc,vdwgg,vxc}

Here we investigate with density functional theory (DFT)
the effects on the graphite structure and the
energetics and the elastic response when potassium is intercalated.
The final intercalate compound is C$_8$K.
The AM intercalate system is interesting in itself and has been the 
focus of numerous experimental investigations.\cite{wada,zabelPRB82,barnard,palmer,li}
Graphitic systems are also ideal test materials in ongoing theory 
development that aims at improving the description of the nonlocal 
interlayer bonds in sparse systems.~\cite{henrik,svetla,girifalco}
Standard DFT
approaches are based on local (local density approximation, LDA) and
semilocal approximations (generalized gradient approximation,
GGA)~\cite{dft1,dft2,dft3,dft4} for the electron exchange and
correlation.  Such regular DFT tools do not treat correctly the 
weak vdW binding, 
e.g., the cohesion between (adjacent) graphite layers. 
The failure of traditional DFT for graphite makes it impossible
to obtain a meaningful comparison of the energetics in on-surface
AM adsorption and subsurface AM absorption. Conversely, investigations
of graphitic systems like C$_8$K permit us to test the accuracy of our
vdW-DF development work.

We explore the nature of the bonding of graphite, the
process leading to intercalation via adsorption and absorption of
potassium, and the nature of 
potassium-intercalated graphite C$_8$K using a recently developed  
vdW-DF density functional.\cite{vdwgg}
This choice of functional is essential for a comparison of graphite and 
C$_8$K properties because of the inability of traditional GGA-based DFT 
to describe graphite. We calculate
the structure  and elastic response (bulk modulus $B_0$) 
of pristine graphite and potassium intercalated graphite
and we present results for the formation energies of the C$_8$K system. 

The intercalation of potassium in graphite is preceded by the
adsorption of potassium on top of a graphite surface and 
potassium absorption underneath the top graphite layer of the 
surface. In this work we study how potassium bonds to graphite in
these two parts of the process towards intercalation.
Our vdW-DF investigations of the binding of potassium in or on graphite supplements
corresponding vdW-DF studies of the binding of  
polycyclic aromatic hydrocarbon dimers, of the polyethylene crystal,
of benzene dimers, 
and of polycyclic aromatic hydrocarbon and phenol molecules on 
graphite.\cite{pahsvetla,PE,bonog,phenol,langreth1,langreth2}

The outline of the paper is as follows. 
Section II contains a short description of the materials of interest here: graphite,
C$_8$K, and graphite with an adsorbed or absorbed K atom layer.
The vdW-DF scheme is described in Sec.~III.
 Section IV presents our results, Sec.~V the 
discussion, and conclusions are drawn in Sec.~VI.
\section{Material structure}

Graphite is a semimetallic solid with strong intra-plane bonds and
weakly coupled layers. The presence of these two types of bonding 
results in a material with different properties along the various
crystallographic directions.\cite{chung} For example, the thermal
and electrical conductivity along the carbon sheets is two orders
of magnitude higher than that perpendicular to the sheets. This
specific property allows heat to move directionally, which makes
it possible to control the heat transfer. 
The relatively weak 
vdW forces between the sheets contribute to another
industrially important property: graphite is an ideal lubricant.
In addition, the anisotropic properties of graphite make the
material suitable as a substrate in electronic studies of
ultrathin metal
films.~\cite{mackan_prb64,hu,jinhe,osterlund}

\begin{figure}
\begin{center}
\scalebox{0.50}{\includegraphics{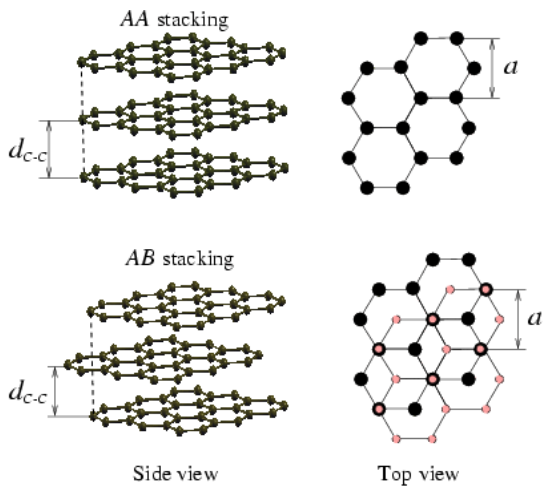}}
\caption{\label{Graphite} (Color online) Simple hexagonal graphite ($AA$
stacking) and natural hexagonal graphite ($AB$ stacking).
The two structures differ by that each
second carbon layer in  $AB$-stacked graphite is shifted,
whereas in  $AA$-stacked graphite all planes are directly
above each other.  The experimentally obtained
in-plane lattice constant and sheet separation of natural graphite
is (Ref.~\onlinecite{baskin})  $a=2.459$\,{\AA} and 
$d_{\mbox{\scriptsize C-C}}=3.336$\,{\AA}, respectively.}
\end{center}
\end{figure}

The natural structure of graphite is an $AB$ stacking,
with the graphite layers shifted relative to each other, as illustrated
in Fig.~\ref{Graphite}. The figure also shows hexagonal graphite, 
consisting of $AA$-stacked graphite layers.
The in-plane lattice constant $a$ and the layer separation 
$d_{\mbox{\scriptsize C-C}}$ is 
also illustrated. In natural graphite the primitive unit cell 
is hexagonal, includes four carbon atoms in two layers, 
and has unit cell side lengths $a$ and height
$c=2d_{\mbox{\scriptsize C-C}}$.

The physical properties of graphite have been studied 
in a variety of experimental\cite{baskin,eberhardt,law} and
theoretical\cite{ahuja1,holzwarth} work.
Some of the DFT work has been performed in LDA,
which does not  provide a physically meaningful
 account of binding in layered systems.\cite{hardnumbers,ijqc}
At the same time, using GGA is not an option because it does not bind 
the graphite layers. For a good description of the graphite
structure and nature 
the vdW interactions must be included.\cite{hardnumbers}

Alkali metals (AM), except Na, easily
penetrate the gallery of the graphite forming alkali metal
graphite intercalation compounds.
 These intercalation
compounds are formed through electron exchange between the
intercalated layer and the host carbon layers, resulting in 
a different nature of the interlayer bonding type than that of pristine
graphite. The intercalate also affects the conductive properties
of graphite, which becomes superconductive in the direction
parallel to the planes at critical temperatures below 1
K.\cite{supercond,jishi}

\begin{figure}
\begin{center}
\scalebox{0.45}{\includegraphics{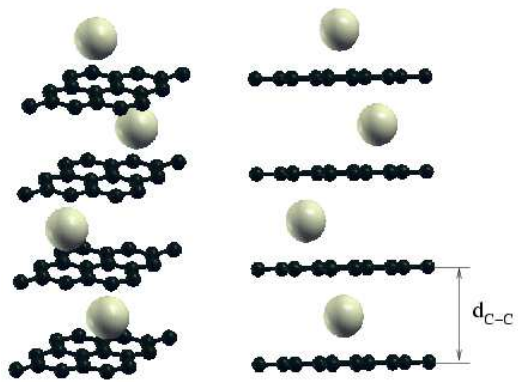}}
\caption{\label{AAAA} (Color online) Crystalline structure of C$_8$K showing the
$AA$-stacking of the carbon layers (small balls) and the
$\alpha\beta\gamma\delta$-stacking of the potassium layers (large
balls) perpendicular to the graphene sheets.
The potassium layers are arranged in a $p(2\times 2)$
structure, with the K atoms occupying the sites over the hollows
of every fourth carbon hexagon.}
\end{center}
\end{figure}

The structure of AM graphite intercalation compounds is characterized by
its stage $n$, where  $n$ is the number of graphite sheets located between
the AM layers.  In this work we consider only stage-1 intercalated graphite
C$_8$K, in which the layers of graphite and potassium alternate throughout
the crystal. The primitive unit cell of C$_8$K
is orthorhombic and contains sixteen C atoms and two K atoms.
In the C$_8$K crystal the K atoms are
ordered in a $p(2\times 2)$ registry
with K-K separation 2$a$, where $a$ is the in-plane lattice
constant of graphite. 
This separation of the potassium atoms is about 8\%
larger than that in the natural K bcc crystal
(based on experimental values).
The carbon sheet stacking in C$_8$K is
of $AA$ type, with the K atoms occupying the sites over the hollows
of every fourth carbon hexagon, each position denoted by $\alpha$, 
$\beta$, $\gamma$, or $\delta$, and
the stacking of the K atoms
perpendicular to the planes being described by the 
$\alpha\beta\gamma\delta$-sequence as illustrated in Fig.~\ref{AAAA}. 

\section{Computational methods}
\label{sec:method}

The first-principle total-energy and electronic structure
calculations are performed within the framework of DFT. 
The  semilocal Perdew-Burke-Ernzerhof
(PBE) flavor\cite{dft2} of 
GGA is chosen for the exchange-correlation functional  for the
traditional self-consistent calculations underlying the vdW-DF calculations. 
For all GGA calculations
we use the open source DFT code \textsc{Dacapo},\cite{dacapo}
which employs Vanderbilt ultrasoft pseudopotentials,\cite{vanderbilt}
periodic boundary conditions, and a plane-wave basis set. An energy
cut-off of 500\,eV is used for the expansion of the wave functions and
the Brillouin zone (BZ) of the unit cells is sampled according to the
Monkhorst-Pack scheme.\cite{monkhorst} The self-consistently determined
GGA valence electron density $n(\mbox{\boldmath $r$})$ as well
as components of the energy from these calculations are passed on to the
subsequent vdW-DF calculation of the total energy.

For the adsorption and absorption studies a graphite surface slab
consisting of 4 layers is used, with a surface unit cell of side lengths
twice those in the graphite bulk unit cell (i.e., side lengths $2a$).
The surface calculations are performed with a
4$\times$4$\times$1 $k$-point sampling of the BZ.

The (pure) graphite bulk GGA calculations are 
performed with a 8$\times$8$\times$4 $k$-point sampling of the BZ,
whereas for the C$_8$K bulk structure,
in a unit cell at least double the size in any direction,  
4$\times$4$\times$2 $k$-points are used, consistent also with
the choice of $k$-point sampling of the surface slabs.

We choose to describe C$_8$K by using a hexagonal unit cell
with four formula units, lateral side lengths approximately twice those of 
graphite and with four graphite and four K-layers in 
the direction perpendicular to the layers.
C$_8$K can also be described by the previously mentioned primitive
orthorhombic unit cell containing two formula units of atoms but we
retain the orthorhombic cell for ease of 
description and for simple implementation of numerically robust vdW-DF calculations.

In all our studies, except test cases, the 
Fast Fourier Transform (FFT) grids are chosen such that the
separation of neighboring points is maximum $\sim$0.13\,{\AA} in any 
direction in any calculation.

\subsection{vdW density function calculations}

In graphite, the carbon layers bind by vdW interactions only. In
the intercalated compound a major part of the attraction is ionic,
but also here the vdW interactions cannot be ignored. In
order to include the vdW interactions systematically in all of our calculations
we use the vdW-DF of Ref.~\onlinecite{vdwgg}. There, the correlation
energy functional is divided into a local and a nonlocal part,
\begin{equation}
E_\mathrm{c}  \approx
E_\mathrm{c}^{\mathrm{LDA}}+E_\mathrm{c}^{\mathrm{nl}},
\label{eq:gg-vdW}
\end{equation}
where the local part is approximated in the LDA and the 
nonlocal part $E_\mathrm{c}^\mathrm{nl}$ is
consistently constructed to vanish for a homogeneous system.  The
nonlocal correlation $E_\mathrm{c}^\mathrm{nl}$ is calculated from
the GGA-based $n(\mbox{\boldmath $r$})$ and its gradients by using
information about the many-body response of the weakly inhomogeneous
electron gas:
\begin{equation}
E_\mathrm{c}^\mathrm{nl}=\frac{1}{2}\int_{V_0} d\mbox{\boldmath $r$}
\int_{V} d\mbox{\boldmath $r$}' n(\mbox{\boldmath $r$})
\phi(\mbox{\boldmath $r$}, \mbox{\boldmath $r$}') n(\mbox{\boldmath
$r$}'). \label{eq:nonlocal2}
\end{equation}
The nonlocal kernel $\phi(\mbox{\boldmath $r$},\mbox{\boldmath
$r$}')$ can be tabulated in terms of the separation
$|\mbox{\boldmath $r$}-\mbox{\boldmath $r$}'|$ between the two
fragments at positions $\mbox{\boldmath $r$}$ and $\mbox{\boldmath
$r$}'$ through the parameters
$D=(q_0+q_0')|\mbox{\boldmath $r$}-\mbox{\boldmath $r$}'|/2$ and
$\delta=(q_0-q'_0)/(q_0+q'_0)$. Here $q_0$ is a local parameter
that depends on the electron density and its gradient at position
$\mbox{\boldmath $r$}$. The analytic expression for the kernel $\phi$ in
terms of $D$ and $\delta$ can be found in Ref.~\onlinecite{vdwgg}.

For periodic systems, such as bulk graphite, C$_8$K, and the
graphite surface (with adsorbed or absorbed K-atoms), the nonlocal 
correlation per unit cell is simply
evaluated from the interaction of the points in the unit cell $V_0$
with points everywhere in space ($V$) in the three (for bulk
graphite and C$_8$K) or two (for the graphite surface) dimensions of
periodicity. Thus, the $V$-integral in
Eq.~(\ref{eq:nonlocal2}) in principle requires a representation of
the electron density infinitely repeated in space. 
In practice, the nonlocal
correlation rapidly converges\cite{bonog} and it suffices with
repetitions of the unit cell a few times in each spatial direction.
For graphite bulk 
the $V$-integral 
is converged when we use a $V$ that
extends 9 (7) times the
original unit cell in directions parallel (perpendicular)
to the sheets. For the potassium investigation a significantly
larger original unit cell is adopted (see Fig.~\ref{AAAA});  
here a fully converged $V$
corresponds to a cell extending five (three) times the
original cell in the direction parallel (perpendicular)
to the sheets for C$_8$K bulk.
To describe the nonlocal correlations (\ref{eq:nonlocal2}) for the 
graphite surface a sufficient $V$ extends
five times the original unit cell along the carbon sheets.

For the exchange energy $E_{\rm x}$ we follow the choice of 
Ref.~\onlinecite{vdwgg} of using revPBE\cite{dft3} exchange.
Among the functionals that we
have easy access to, the revPBE has proved to be the best candidate for
minimizing the tendency of artificial exchange binding in graphite.\cite{ijqc} 

Using the scheme described above to evaluate 
$E_{\rm c}^{\rm nl}$, the total energy finally reads:
\begin{equation}
E^{\rm vdW-DF} = E^{\rm GGA}
-E_{\rm c}^{\rm GGA}+E_{\rm c}^{\rm LDA}+E_{\rm c}^{\rm nl},
\label{eq:vdW}
\end{equation}
where $E^{\rm GGA}$ is the GGA total energy with the revPBE choice
for the exchange description and $E_{\rm c}^{\rm GGA}$ ($E_{\rm
c}^{\rm LDA}$) the GGA (LDA) correlation energy.
As our GGA calculations in this specific application
of vdW-DF are carried out in PBE, not revPBE, we further need to explicitly
replace the PBE exchange in $E^{\rm GGA}$ by that of revPBE
for the same electron charge density distribution.

\subsection{Convergence of the local and nonlocal energy variation}

DFT calculations provide physically meaningful results for energy differences
between total energies (\ref{eq:vdW}). To understand  materials and processes we
must compare total energy differences between
a system with all constituents at relatively close
distance and a system of two or more fragments at ``infinite" separation
(the reference system).
Since the total energy (\ref{eq:vdW}) consists both of a long-range
term  and shorter-ranged GGA and LDA terms it is natural to choose
different ways to represent the separated fragments for these different 
long- or short-range energy terms. 

For the shorter-range energy parts (LDA and GGA terms) the reference 
system is a full system with vacuum between the fragments. For 
LDA and GGA  calculations it normally suffices to make sure that
the charge density tails of the fragments do not overlap, but
here we find that the surface dipoles cause a slower 
convergence with layer separation.  We use a system with the 
layer separation between the potassium layer and the nearest 
graphite layer(s) $d_{\mbox{\scriptsize C-K}}=12$ {\AA} 
(8 {\AA}) as reference for the adsorption (absorption) study.
 
The evaluation of the nonlocal correlations $E_\mathrm{c}^\mathrm{nl}$
requires additional care. This is due to technical reasons pertaining 
to numerically stability in basing the $E_\mathrm{c}^\mathrm{nl}$ 
evaluation on the FFT grid used to converge the underlying 
traditional-DFT calculations.  The evaluation of the nonlocal 
correlation energy, Eq.~(\ref{eq:nonlocal2}), involves a weighted double integral 
of a kernel with a significant short-range variation \cite{vdwgg}. 
The shape of the kernel makes the $E_\mathrm{c}^\mathrm{nl}$
evaluation  sensitive to the particulars of FFT-type griding,\cite{JKPH}
for example, to the relative position of FFT grid points relative 
to the nuclei position (for a finite grid-point spacing). 

However, robust evaluation of binding- or cohesive-energy contributions 
by nonlocal correlations can generally be secured by a further 
splitting of energy differences into steps that minimize
 the above-mentioned grid sensitivity. The problem of FFT 
sensitivity of the $E_\mathrm{c}^\mathrm{nl}$ evaluation 
is accentuated because the binding in the $E_\mathrm{c}^\mathrm{nl}$ 
channel arises as a smaller energy difference between sizable 
$E_\mathrm{c}^\mathrm{nl}$ contributions of the system and of the 
fragments. Conversely, convergence in vdW-DF calculations
of binding and cohesive energies can be obtained even at 
a moderate FFT grid accuracy (0.13 {\AA} used here) by devising 
a calculational scheme that always maintains identical position 
of the nuclei relative to grid points in the combined systems 
as well as in the fragment reference system.

Thus we obtain a numerically robust evaluation of the $E_{\rm c}^{\rm nl}$ 
energy differences by choosing steps for which we can explicitly
control the FFT griding. For adsorption and absorption cases
we calculate the reference systems as a sum of 
$E_{\rm c}^{\rm nl}$-contributions for each fragment and 
we make sure to always position the fragment at the exact same
position in the system as in the interacting system. For bulk systems
we choose steps in which we exclusively adjust the inter-plane or 
in-plane lattice constant. Here the reference system is then
simply defined as a system with either double (or in some cases
quadruple) lattice constant and with a corresponding doubling
of the FFT griding along the relevant unit-cell vector.

The cost of full convergence is that, in practice, we often do three or more GGA 
calculations and subsequent $E_{\rm c}^{\rm nl}$ calculations for 
each point on the absorption, absorption, or formation-energy 
curve. In addition to the calculations for the full system we have
to do one for each of the isolated fragments at identical position 
in the adsorption/absorption cases and one or more for fragments in
the doubled unit-cell and doubled griding reference. 
We have explicitly tested that using a FFT grid spacing of $< 0.13$ {\AA}
(but not larger) for such reference calculations is sufficient to
ensure full convergence in the reported $E_{\rm c}^{\rm nl}$ (and $E^{\rm vdW-DF}$
total) energy variation for graphitic systems. 
 
\subsection{Material formation and sorption energies}

The cohesive energy of graphite (G) is the energy gain, per carbon
atom, of creating
graphite at in-plane lattice constant $a$ and layer separation 
$d_{\mbox{\scriptsize C-C}}$ from isolated (spin-polarized) carbon atoms.
\begin{equation}
E_{\mbox{\scriptsize G,coh}}(a,d_{\mbox{\scriptsize C-C}})= 
E_{\mbox{\scriptsize G,tot}}(a,d_{\mbox{\scriptsize C-C}}) - 
E_{\mbox{\scriptsize C-atom,tot}}
\end{equation} 
where $E_{\mbox{\scriptsize G,tot}}$ and $E_{\mbox{\scriptsize C-atom,tot}}$ are
total energies per carbon atom.
The graphite structure is stable at the minimum of the cohesive energy, at
lattice constants $a=a_{\mbox{\scriptsize G}}$ and 
$2d_{\mbox{\scriptsize C-C}}=c_{\mbox{\scriptsize G}}$.

The adsorption (absorption) energy for a $p(2\times 2)$ K-layer over (under) the
top layer of a graphite surface is the difference in total 
energy [from Eq.\ (\ref{eq:vdW})] for the system at hand minus 
the total energy of the initial system, i.e., a clean graphite 
surface and isolated gas-phase potassium atoms. However, due 
to the above mentioned technical issues in using the vdW-DF we calculate the 
adsorption and absorption energy as a sum of (artificial) stages leading
to the desired system: First the initially isolated, spin-polarized potassium
atoms are gathered into a free floating potassium 
layer with the structure corresponding to a full cover of potassium atoms. 
By this the total system gains the energy 
$\Delta E_{\mbox{\scriptsize K-layer}}(a_{\mbox{\scriptsize G}})$,
with
\begin{equation}
\Delta E_{\mbox{\scriptsize K-layer}}(a)
=E_{\mbox{\scriptsize K,tot}}(a)
-E_{\mbox{\scriptsize K-atom,tot}}
\,.
\end{equation}

In \textit{adsorption\/} the potassium layer is then simply placed on top 
of the four-layer $(2\times 2)$ graphite surface (with the K atoms above 
graphite hollows) at distance $d_{\mbox{\scriptsize C-K}}$. 
The system thereby gains a further energy contribution 
$\Delta E_{\mbox{\scriptsize K-G}}(d_{\mbox{\scriptsize C-K}})$.
This leads to an adsorption energy per K-atom
\begin{equation}
E_{\mbox{\scriptsize ads}}(d_{\mbox{\scriptsize C-K}})=
\Delta E_{\mbox{\scriptsize K-layer}}(a_{\mbox{\scriptsize G}})
+\Delta E_{\mbox{\scriptsize K-G}}(d_{\mbox{\scriptsize C-K}})\,.
\label{eq:ads}
\end{equation} 

In \textit{absorption\/} the 
top graphite layer is peeled off the $(2\times 2)$ graphite surface
and moved to a distance far from the remains of  
the graphite surface. This process costs the system an (``exfoliation'') 
energy $-\Delta E_{\mbox{\scriptsize C-G}}=
-[E_{\mbox{\scriptsize tot,C-G}}(d_{\mbox{\scriptsize C-C}}=c_{\mbox{\scriptsize G}}/2)
-E_{\mbox{\scriptsize tot,C-G}}(d_{\mbox{\scriptsize C-C}}\rightarrow\infty)]$. 
At the far distance the 
isolated graphite layer is moved into $AA$ stacking with the 
surface, at no extra energy cost.
Then, the potassium layer is placed midway between the far-away
graphite layer and the remains of the graphite surface. 
Finally the two layers are gradually moved towards the 
surface. At distance $2d_{\mbox{\scriptsize C-K}}$ between the two
topmost graphite layers (sandwiching the K-layer) 
the system has further gained an energy 
$\Delta E_{\mbox{\scriptsize C-K-G}}(d_{\mbox{\scriptsize C-K}})$.
The absorption energy per K-atom is thus
\begin{equation}
E_{\mbox{\scriptsize abs}}(d_{\mbox{\scriptsize C-K}})=
{} -\Delta E_{\mbox{\scriptsize C-G}} 
+ \Delta E_{\mbox{\scriptsize K-layer}}(a_{\mbox{\scriptsize G}})
+ \Delta E_{\mbox{\scriptsize C-K-G}}(d_{\mbox{\scriptsize C-K}}) \,.
\label{eq:abs}
\end{equation} 

Similarly, the C$_8$K intercalate compound is formed from graphite 
by first moving 
the graphite layers far apart accordion-like (and there shift 
the graphite stacking from $ABA\ldots$ to $AAA\ldots$ at no energy 
cost), then changing the in-plane lattice constant of the isolated
graphene layers from $a_{\mbox{\scriptsize G}}$ to $a$,
then intercalating K-layers (in stacking $\alpha\beta\gamma\delta$) 
between the graphite layers, and finally moving all the K- and graphite 
layers back like an accordion, with in-plane lattice constant $a$
(which has the value $a_{\mbox{\scriptsize C$_8$K}}$ at equilibrium).

In practice, a unit cell of four periodically
repeated graphite layers is used in order to accommodate the potassium
$\alpha\beta\gamma\delta$-stacking. 
The energy gain of creating a $(2\times 2)$ graphene sheet from 8 
isolated carbon atoms is defined similarly to that of the K-layer:
\begin{equation}
\Delta E_{\mbox{\scriptsize C-layer}}(a)=E_{\mbox{\scriptsize C-layer,tot}}(a)
-8E_{\mbox{\scriptsize C-atom,tot}} \,.
\end{equation}

The formation energy for the C$_8$K intercalate compound
per K atom or formula unit, $E_{\mbox{\scriptsize form}}$, 
is thus found from
the energy cost of moving four graphite layers apart by expanding the
$(2\times 2)$ unit cell to large height, $-\Delta E_{\mbox{\scriptsize G-acc}}$, 
the cost of changing the in-plane lattice constant from 
$a_{\mbox{\scriptsize G}}$ to $a$ in each of the four isolated graphene layers,
$4(\Delta E_{\mbox{\scriptsize C-layer}}(a)-\Delta E_{\mbox{\scriptsize C-layer}}(a_{\mbox{\scriptsize G}}))$,
the gain of creating four K-layers from isolated K-atoms, 
$4\Delta E_{\mbox{\scriptsize K-layer}}(a)$, plus 
the gain of bringing four K-layers and four graphite layers together in 
the C$_8$K structure,
$\Delta E_{\mbox{\scriptsize C$_8$K-acc}}(a,d_{\mbox{\scriptsize C-K}})$,
yielding
\begin{eqnarray}
\lefteqn{E_{\mbox{\scriptsize form}}(a,d_{\mbox{\scriptsize C-K}})}
\nonumber\\[0.6em]
&=& {\textstyle \frac{1}{4}} \Big[
-\Delta E_{\mbox{\scriptsize G-acc}} 
+4\Delta E_{\mbox{\scriptsize C-layer}}(a)
-4\Delta E_{\mbox{\scriptsize C-layer}}(a_{\mbox{\scriptsize G}})
\nonumber \\[0.6em]
&&{}
+4\Delta E_{\mbox{\scriptsize K-layer}}(a)
+\Delta E_{\mbox{\scriptsize C$_8$K-acc}}(a,d_{\mbox{\scriptsize C-K}}) \Big]\,.
\label{eq:form}
\end{eqnarray} 

The relevant energies to use for comparing the three different
mechanisms of including potassium (adsorption, absorption and intercalation)
are thus $E_{\mbox{\scriptsize ads}}(d_{\mbox{\scriptsize C-K}})$,
 $E_{\mbox{\scriptsize abs}}(d_{\mbox{\scriptsize C-K}})$  and
$E_{\mbox{\scriptsize form}}(a,d_{\mbox{\scriptsize C-K}})$ at their
respective minimum values.

\section{Results}

\label{sec:results}

Experimental observations indicate that the intercalation of 
potassium into graphite starts with the \textit{ab}sorption of
evaporated potassium into an initially clean graphite surface.\cite{footnote}
This subsurface absorption is preceded by initial, sparse potassium 
\textit{ad}sorption onto the
surface,
and proceeds with further absorption into deeper graphite voids.
The general view is that the K atoms enter graphite at the 
graphite step edges.\cite{barnard}
The amount and position of intercalated K atoms is controlled by the
temperature and time of evaporation.

Below, we first describe the initial clean graphite system,
and the energy gain in (artificially) creating free-floating
K-layers from isolated K-atoms. 
Then we present and discuss our results on  
potassium adsorption and subsurface absorption, followed
by a characterization of bulk C$_8$K.

For the adsorption (absorption) system we calculate the adsorption
(absorption)
energy curve, including the equilibrium structure. As  
a demonstration of the need for a relatively fine FFT griding in 
the vdW-DF calculations we also calculate and compare 
the absorption curve for a more sparse FFT grid. 
For the bulk systems (graphite and C$_8$K) we determine the 
lattice parameters and
the bulk modulus. 
We also calculate the 
formation energy of C$_8$K and the energy needed to peel off 
one graphite layer from the graphite surface and compare
with experiment.\cite{zacharia}

\subsection{Graphite bulk structure}
The present calculations on pure graphite are for the natural, 
$AB$-stacked graphite (lower panel of Fig.~\ref{Graphite}).  The cohesive energy is calculated
at a total of 
232 structure values $(a,\,d_{\mbox{\scriptsize C-C}})$ and the
equilibrium structure and bulk modulus $B_0$ 
are then evaluated using the method described in
Ref.~\onlinecite{bulkmodulus}.

\begin{figure}
\begin{center}
\scalebox{0.65}{\includegraphics{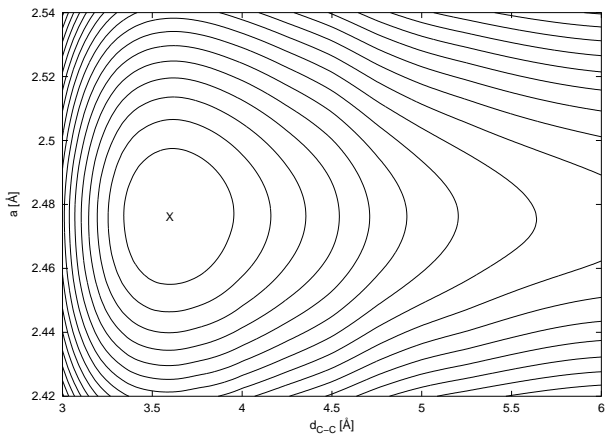}}
\caption{\label{Fig_graphite} Graphite cohesive energy
$E_{\mbox{\scriptsize G,coh}}$ ($AB$-stacked), based on vdW-DF, as a function 
of the carbon layer separation
$d_{\mbox{\scriptsize C-C}}$ and the in-plane lattice constant $a$.
The energy contours are spaced by 5\,meV per carbon atom.}
\end{center}
\end{figure}

Figure~\ref{Fig_graphite} shows a contour plot of the graphite
cohesive energy variation $E_{\mbox{\scriptsize G,coh}}$ 
as a function of the layer separation
$d_{\mbox{\scriptsize C-C}}$ and the in-plane lattice constant $a$,
calculated within the vdW-DF scheme. The contour spacing 
is 5\,meV per carbon atom, shown relative to the energy minimum 
located at $(a$, $d_{\mbox{\scriptsize C-C}})
=(a_{\mbox{\scriptsize G}}$, $c_{\mbox{\scriptsize G}}/2)
=$(2.476\,{\AA}, 3.59\,{\AA}).
These values are summarized in Table~\ref{tableK} together with the results
obtained from a semilocal PBE calculation. 
As expected, and discussed in Ref.~\onlinecite{henrik}, the semilocal PBE
calculation yields unrealistic results for the layer separation.
The table also presents the corresponding experimental values.
Our calculated lattice values obtained using vdW-DFT are in good
agreement with experiment,\cite{baskin} 
and close to those found from the older vdW-DF of 
Refs.~\onlinecite{henrik} and~\onlinecite{ijqc},
(in which we for 
$E_{\rm c}^{\rm nl}$
assume translational invariance of $n(\mbox{\boldmath $r$})$
along the graphite planes,)  at (2.47\,{\AA}, 3.76\,{\AA}).

Consistent with experimental reports\cite{wada} and our previous 
calculations\cite{hardnumbers,ijqc,henrik} 
we find graphite to be rather soft, indicated by the
bulk modulus $B_0$ value. Since in-plane compression is very hard
in graphite most of the softness suggested by (the isotropic)
 $B_0$ comes from compression perpendicular to the graphite layers,
and the value of $B_0$ is expected to be almost identical to the
$C_{33}$ elastic coefficient.\cite{wada,henrik}

\begin{table}
\begin{center}
\caption{\label{tableK} Optimized structure parameters and elastic
properties for natural hexagonal graphite ($AB$-stacking) and
the potassium-intercalated graphite structure C$_8$K in
$A\alpha A \beta A\gamma A\delta A \alpha\ldots$ stacking.
The table shows the calculated optimal values of the in-plane 
lattice constant $a$, the (graphite-)layer-layer separation 
$d_{\mbox{\scriptsize C-C}}$, and the bulk modulus
$B_0$. 
In C$_8$K the value if $d_{\mbox{\scriptsize C-C}}$ is twice the 
graphite-potassium distance $d_{\mbox{\scriptsize C-K}}$.}

\vspace{1em}

\begin{tabular}{lccccccc}
&\multicolumn{3}{c}{Graphite} &\hspace{1.5ex}&\multicolumn{3}{c}{C$_8$K}  \\
\cline{2-4}\cline{6-8}
& PBE& vdW-DF & Exp.  & &PBE & vdW-DF & Exp.\\
\hline
$a$ (\AA) &2.473 &2.476 & 2.459$^a$ & & 2.494 & 2.494 & 2.480$^b$\\
$d_{\mbox{\scriptsize C-C}}$ (\AA) & $\gg 4$ &3.59 & 3.336$^a$  && 5.39 & 5.53  & 5.35$^c$\\
$B_0$ (GPa) & &27 &$37^{de}$ & & 37 & 26 & 47$^{de}$ 
\\
\hline
\multicolumn{8}{l}{%
$^a$Ref.~\protect\onlinecite{baskin}.
$^b$Ref.~\protect\onlinecite{nixon1969}.
$^c$Ref.~\protect\onlinecite{k_intercalation}.
$^d$Ref.~\protect\onlinecite{wada}.%
}\\
\multicolumn{8}{l}{$^e$Value presented is for $C_{33}$; 
for laterally rigid materials, like} \\
\multicolumn{8}{l}{graphite and C$_8$K, $C_{33}$
is a good approximation of $B_0$.}
\end{tabular}
\end{center}
\end{table}

We find the energy cost of peeling off a graphite layer
from the graphite surface  (the exfoliation energy) to be 
$\Delta E_{\mbox{\scriptsize C-G}}= -435$ meV per  
($2\times 2$) unit cell, i.e., $-55$ meV per surface carbon atom 
(Table \ref{table_formation}).
A recent experiment\cite{zacharia} measured the 
desorption energy of polycyclic aromatic hydrocarbons
(basically flakes of graphite sheets) off a graphite surface. From
this experiment the energy cost of peeling off a graphite layer 
from the graphite surface was deduced to $-52\pm 5$\,meV/atom. 

Our value $-55$ meV/C-atom is also consistent with a separate vdW-DF
determination\cite{pahsvetla} of the binding ($-47$\,meV per in-plane atom)
between two (otherwise) isolated graphene sheets. 

For the energies of the absorbate system and of the C$_8$K 
intercalate a few other graphite-related energy contributions are needed.
The energy of collecting C atoms to form a graphene sheet at
lattice constant $a$ from isolated (spinpolarized) atoms is
given by $\Delta E_{\mbox{\scriptsize C-layer}}(a)$; 
we find that changing the lattice constant $a$ from  $a_{\mbox{\scriptsize G}}$ 
to the equilibrium value $a_{\mbox{\scriptsize C$_8$K}}$
of C$_8$K causes this energy to change
a mere 30 meV per ($2\times 2$) sheet.
The contribution $\Delta E_{\mbox{\scriptsize G-acc}}$ is the energy 
of moving bulk graphite layers (in this case four periodically repeated 
layers) far away from each other, by expanding
the unit cell along the direction perpendicular to the layers.
Thus, $\Delta E_{\mbox{\scriptsize G-acc}}
=32\Delta E_{\mbox{\scriptsize G,coh}}(a_{\mbox{\scriptsize G}},c_{\mbox{\scriptsize G}}/2)
-4\Delta E_{\mbox{\scriptsize C-layer}}(a_{\mbox{\scriptsize G}})$ 
taking the number of atoms and layers per unit cell into account. 
We find the value $\Delta E_{\mbox{\scriptsize G-acc}}= -1600$ meV per 
($2\times 2$) four-layer unit cell. This corresponds to $-50$ meV per C atom, 
again consistent with our result for the exfoliation energy, 
$\Delta E_{\mbox{\scriptsize C-G}}/8 = -55$ meV.

\subsection{Creating a layer of K-atoms}

The (artificial) step of creating a layer of 
potassium atoms from isolated atoms releases a significant energy 
$\Delta E_{\mbox{\scriptsize K-layer}}$. This energy
contains the  
energy variation with in-plane lattice constant and the energy cost 
of changing from a spin-polarized to a spin-balanced electron configuration
for the isolated atom.\cite{wilkins}

The creation of the K-layer provides an energy gain 
which is about half an eV per potassium atom, depending on the final lattice
constant. With the graphite lattice constant $a_{\mbox{\scriptsize G}}$
the energy change, including the spin-change cost, is 
$\Delta E_{\mbox{\scriptsize K-layer}}(a_{\mbox{\scriptsize G}}) = -476$\,meV 
per K atom in vdW-DF ($-624$\,meV when calculated within PBE), whereas 
$\Delta E_{\mbox{\scriptsize K-layer}}(a_{\mbox{\scriptsize C$_8$K}}) = -473$\,meV
in vdW-DF.

\subsection{Graphite-on-surface adsorption of potassium}

The potassium atoms are adsorbed on a usual $ABA\ldots$-stacked graphite
surface. We consider here full (one monolayer) coverage, which is one
potassium atom per ($2\times 2$) graphite surface unit cell. This orders
the potassium atoms in a honeycomb structure with lattice constant
2$a_{\mbox{\scriptsize G}}$, and a nearest-neighbor distance within the
K-layer of $a_{\mbox{\scriptsize G}}$. 

The unit cell used in the standard DFT calculations 
for adsorption and absorption
has a height of 40\,{\AA} and
includes a vacuum region sufficiently big that no interactions (within GGA)
can occur between the top graphene sheet and the slab bottom in the periodically
repeated image of the slab. The vacuum region is also large in order to guarantee
that the separation from any atom to the dipole layer\cite{bengtsson} always
remains larger than 4 {\AA}.

\begin{figure}
\begin{center}
\scalebox{1.0}{\includegraphics{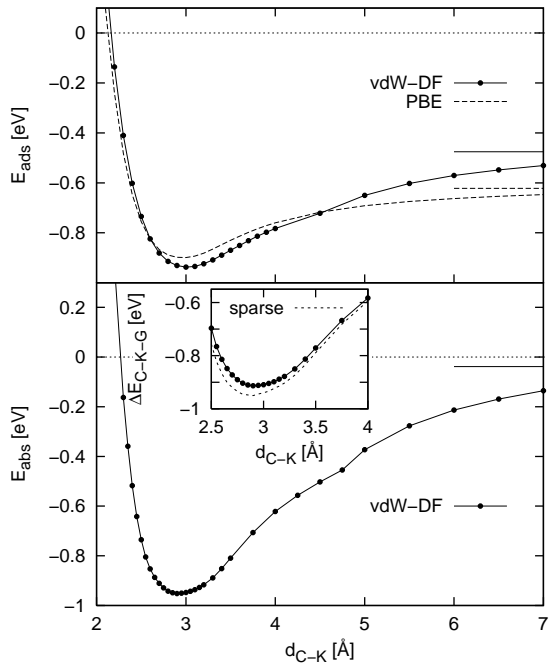}}
\caption{\label{Surface} Potassium adsorption and absorption energy 
at the graphite surface as a function of
the separation $d_{\mbox{\scriptsize C-K}}$ of the K-atom layer
and the nearest graphite layer(s) (at in-plane lattice constant
corresponding to that of the surface, $a_{\mbox{\scriptsize G}}$).
\textit{Top panel:\/} Adsorption curve
based on vdW-DF calculations (solid line with black circles) and 
PBE GGA calculations (dashed line). The horizontal lines to
the left show the energy gain in creating the isolated K layer 
from isolated atoms, 
$\Delta E_{\mbox{\scriptsize K-layer}}(a_{\mbox{\scriptsize G}})$,
the asymptote of  $E_{\mbox{\scriptsize ads}}(d_{\mbox{\scriptsize C-K}})$  
in this plot.
\textit{Bottom panel:\/}
Absorption curve based on vdW-DF calculations. The asymptote is here
the sum $\Delta E_{\mbox{\scriptsize K-layer}}(a_{\mbox{\scriptsize G}})
-\Delta E_{\mbox{\scriptsize C-G}}$.
\textit{Inset:\/} Binding energy of the K-layer and the top graphite layer
(``C-layer'') on top of the graphite slab, $\Delta E_{\mbox{\scriptsize C-K-G}}$.  
The dashed curve shows our results when in $E_{\rm c}^{\rm nl}$ ignoring
every second FFT grid point (in each direction) of
the charge density from the underlying GGA calculations, the solid curve
with black circles shows the result of using every available FFT grid point.}
\end{center}
\end{figure}

In the top panel of Fig.\ \ref{Surface} we show the adsorption energy 
per potassium atom.
The adsorption energy at equilibrium is $-937$ meV per K atom at distance
$d_{\mbox{\scriptsize C-K}}=3.02$ {\AA} from the graphite surface.

For comparison we also show the adsorption curve calculated in a
PBE-only traditional DFT calculation.  Since the interaction
between the K-layer and the graphite surface has a short-range 
component to it, even GGA calculations, such as the PBE curve, show
significant binding ($-900$ meV/K-atom at $d_{\mbox{\scriptsize C-K}}=2.96$ {\AA}). 
This is in contrast to the pure vdW binding between
the layers in clean graphite.\cite{ijqc,henrik} Note that the asymptote of
the PBE curve is different from that of the vdW-DF curve, this is due to 
the different energy gains ($\Delta E_{\mbox{\scriptsize K-layer}}$) in
collecting a potassium layer from isolated atoms when calculated in PBE 
or in vdW-DF.

For K-adsorption the vdW-DF and PBE curves agree reasonably well, and
the use of vdW-DF for this specific calculation is not urgently necessary.
However, in order to compare the adsorption results consistently to 
absorption, intercalation and clean graphite, it is necessary to 
include the long-range interactions through vdW-DF. As shown for the 
graphite bulk results above, PBE yields quantitatively and qualitatively
wrong results for the layer separation.

\subsection{Graphite-subsurface absorption of potassium}

The first subsurface adsorption of K takes place in the void under the 
top-most graphite layer. The surface absorption of the
first K-layer causes a lateral shift of the top graphite sheet, resulting in
a $A/$K$/ABAB\ldots$ stacking of the graphite. We have studied the bonding nature of
this absorption process by considering a full $p(2\times 2)$-intercalated 
potassium layer in the subsurface of a four layer thick graphite slab.

Following the receipt of Section \ref{sec:method}
for the absorption energy (\ref{eq:abs}) the energies 
$\Delta E_{\mbox{\scriptsize C-K-G}}$
are approximated by those from a four-layer intercalated 
graphite slab with the stacking $A/$K$/ABA$, and the 
values are shown in the inset of Fig.\ \ref{Surface}. The absorption energy 
$E_{\mbox{\scriptsize abs}}$
is given by the curve in the bottom panel of Fig.\ \ref{Surface}, and its minimum 
is $-952$\,meV per K atom at $d_{\mbox{\scriptsize C-K}}=2.90$ {\AA}. 

To investigate what grid spacing is sufficiently dense to obtain
converged total-energy values in vdW-DF we do additional calculations 
in the binding distance region with a more sparse grid.
Specifically, the inset of Fig.~\ref{Surface} compares the vdW-DF calculated
at full griding with one that uses only every other FFT grid point in each
direction, implying a grid spacing for $E_{\rm c}^{\rm nl}$
(but not for the local terms) which is maximum 0.26\,{\AA}.
We note that
using the full grid yields smaller absolute values of the absorption energy.
We also notice that the effect is more pronounced for small separations
than for larger distances. Thus given resources, the dense FFT grid calculations
are preferred, but even the less dense FFT grid calculations yield reasonably
well-converged results.
In all calculations (except tests of our graphitic systems) we use a  
spacing with maximum 0.13\,{\AA} between grid points. 
This is a grid spacing for which we have explicitly
tested convergence of the vdW-DF for graphitic systems given 
the computational strategy described and discussed in Sec.~III.

\subsection{Potassium-intercalated graphite}

\begin{table}
\begin{center}
\caption{\label{table_formation} Comparison of the graphite exfoliation energy
per surface atom, 
$E_{\mbox{\scriptsize C-G}}/8$, graphite layer binding energy per carbon atom, 
$\Delta E_{\mbox{\scriptsize C-acc}}/32$, 
the energy gain per K atom of 
collecting K- and graphite-layers at equilibrium
to form  C$_8$K, 
$ \Delta E_{\mbox{\scriptsize C$_8$K-acc}}/4$, and the 
equilibrium formation
energy of C$_8$K, $E_{\mbox{\scriptsize form}}$.
}

\vspace{1em}

\begin{tabular}{lcccc}
& $\Delta E_{\mbox{\scriptsize C-G}}/8$ & $\Delta E_{\mbox{\scriptsize C-acc}}/32$ & 
$\Delta E_{\mbox{\scriptsize C$_8$K-acc}}/4$ & 
$E_{\mbox{\scriptsize form}}$\\
& [meV/atom] & [meV/atom] & [meV/${\rm C}_8{\rm K}$] & [meV/${\rm C}_8{\rm K}$]\\
\hline
vdW-DF & $-55$ & $-50$  & $-818$ & $-861$ 
\\ 
PBE & $-$  & $-$ & $-511$ & $-$  
\\ 
Exp. & $-52\pm 5$$^a$& &  & $-1236^b$ \\
\hline 
\multicolumn{4}{l}{%
$^a$Ref.~\protect\onlinecite{zacharia}.
$^b$Ref.~\protect\onlinecite{aronson}.}\\
\end{tabular}
\end{center}
\end{table}

When potassium atoms penetrate the gallery of the graphite, they form planes
that are ordered in a $p(2\times 2)$ fashion along the planes.
The K intercalation causes a shift of every second carbon
layer resulting into an $AA$ stacking of the
graphite sheets. The K atoms then simply occupy the sites over the hollows
of every fourth carbon hexagon. The order of the K atoms perpendicular to
the planes is described by the $\alpha\beta\gamma\delta$ stacking, illustrated
in Fig.~\ref{AAAA}.

For the potassium intercalated compound C$_8$K we calculate in standard
DFT using PBE the total energy at 132 different combinations of the 
structural parameters $a$ and $d_{\mbox{\scriptsize C-C}}$. 
The charge densities and energy terms of these calculations are then
used as input to vdW-DF.
The equilibrium structure and elastic
properties ($B_0$) both for the vdW-DF results and for the PBE
results are then evaluated with the same method as in
the graphite case.~\cite{bulkmodulus} 

\begin{figure}
\begin{center}
\scalebox{1}{\includegraphics{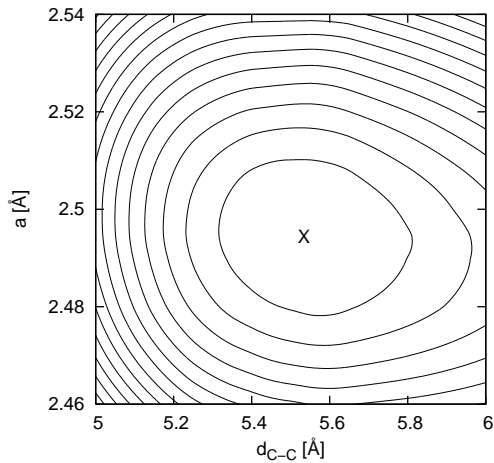}}
\caption{\label{Fig_C8K} 
Formation energy of C$_8$K, $E_{\mbox{\scriptsize form}}$,
as a function of the carbon-to-carbon layer separation
$d_{\mbox{\scriptsize C-C}}$ and half the in-plane lattice constant, $a$. 
The energy contours are spaced by 20\,meV per formula unit.}
\end{center}
\end{figure}

Figure \ref{Fig_C8K} shows a contour plot of the C$_8$K formation 
energy, calculated in vdW-DF, as a function of the C-C layer 
separation ($d_{\mbox{\scriptsize C-C}}$) and the in-plane 
periodicity ($a$) of the graphite-layer structure. The contour 
spacing is 20 meV per formula unit and are shown relative to the 
energy minimum at ($a$, $d_{\mbox{\scriptsize C-C}})=($2.494 {\AA}, 5.53 {\AA}).

\section{Discussion}

Table~\ref{tableK} presents an overview of our structural 
results obtained with the vdW-DF for graphite and C$_8$K. 
The table also contrasts the results with the corresponding 
values calculated with PBE where available.  
The vdW-DF value $d_{\mbox{\scriptsize C-C}}=5.53$\,{\AA} for 
the C$_8$K $C-C$ layer separation is 3\% larger than the 
experimentally observed value whereas the PBE value corresponds 
to less than a 1\% expansion. Our vdW-DF result for the C$_8$K
bulk modulus (26 GPa) is also softer than the PBE result (37 GPa)
and further away from the experimental estimates (47 GPa) based on 
measurements of the $C_{33}$ elastic response.\cite{wada}
A small overestimation of atomic separation is consistent with
the vdW-DF behavior that has been documented in a wide range of 
both finite and extended 
systems.\cite{henrik,ijqc,vdwgg,vxc,pahsvetla,PE,langreth1,langreth2} 
This overestimation results, at least in part, from
our choice of parametrization of the exchange behavior ---
an aspect that lies beyond the present vdW-DF implementation 
which focuses on improving the account of the nonlocal correlations, 
per se.  It is likely that systematic investigations of the
exchange effects can further refine the accuracy of vdW-DF
implementations.\cite{footnote3} In any case, vdW-DF theory
calculations represent, in contrast
to PBE, the only approach to obtain a full ab initio 
characterization of the AM intercalation process.

The C$_8$K system is more compact than graphite and this explains 
why PBE alone can here provide a good description of the materials structure
and at least some materials properties, whereas it fails completely 
for graphite. The distance between the graphene sheets upon 
intercalation of potassium atoms is stretched compared to that 
of pure graphite, but the \mbox{(K-)layer} to (graphite-)layer 
separation, $d_{\mbox{\scriptsize C-K}}=
d_{\mbox{\scriptsize C-C}}/2=2.77$\,{\AA}, 
is significantly less than the layer-layer separation in pure graphite.
This indicates that C$_8$K is likely held together, at least in 
part, by shorter-ranged interactions. 

Table \ref{table_formation} documents that the vdW binding
nevertheless plays an important role in the binding and formation
of C$_8$K. The table summarizes and contrasts our vdW-DF and PBE 
results for graphite exfoliation and layer binding energies as well
as C$_8$K interlayer binding and formation energies.  The
vdW-DF result for the C$_8$K formation energy is smaller than
experimental measurements by 31\% but it nevertheless
represent a physically motivated ab initio calculation.
In contrast, the C$_8$K formation energy is simply  unavailable 
in PBE because PBE, as indicated, fails to describe the layer 
binding in graphite. Moreover, for the vdW-DF/PBE comparisons that 
we can make --- for example, of the C$_8$K layer interaction 
$\Delta E_{\mbox{\scriptsize C$_8$K-acc}}$ --- the vdW-DF is 
found to significantly strengthen the bonding compared with PBE.

It is also interesting to note that the combination of shorter-ranged 
and vdW bonding components in C$_8$K yields a layer binding energy that is
close to that of the graphite case. In spite of the difference in 
nature of interactions, we find almost identical binding energies 
per layer for
the case of the exfoliation and accordion in graphite and for the 
accordion in C$_8$K. This observation testifies to a perhaps 
surprising strength of the so-called soft-matter vdW interactions.

In a wider perspective our vdW-DF permits a first comparison 
of the range of AM-graphite systems from adsorption over absorption 
to full intercalation and thus insight on the intercalation progress.
Assuming a dense $2\times 2$ configuration, we find that the energy for
potassium adsorption and absorption is nearly degenerate with an 
indication that absorption is slightly preferred, consistent with 
experimental behavior.  We also find that the potassium absorption 
may eventually proceeds towards full intercalation thanks to 
a significant release of formation energy. 

\section{Conclusions}

The potassium intercalation process in graphite has been investigated by
means of the vdW-DF  density functional method. This method includes
the dispersive interactions needed for a consistent investigation of
the intercalation process. For clean graphite the vdW-DF
predicts --- contrary to standard semilocal DFT implementations --- a
stabilized bulk system with equilibrium crystal parameters in close
agreement with experiments. Two limits of the absorption process
have been investigated by the vdW-DF, namely single layer subsurface
absorption and the fully potassium intercalated stage-1 crystal C$_8$K. 
Here the vdW-DF is shown to enhance the (semi-)local type of bonding described
by traditional approaches. 
The significant impact on the materials behavior indicates that the 
vdW-DF is needed not only for a consistent
description of sparse matter systems that are solely stabilized by dispersion
forces, but also for their intercalates. 

We thank D.C.\ Langreth and B.I.\ Lundqvist for stimulating discussions.
Partial support from  the Swedish Research Council (VR), the Swedish 
National Graduate School in Materials Science (NFSM), and
the Swedish Foundation for Strategic Research (SSF) through the 
consortium ATOMICS is gratefully acknowledged, 
as well as allocation of computer time at UNICC/C3SE (Chalmers) and 
SNIC (Swedish National Infrastructure for Computing).



\begin{thebibliography}{99}
\bibitem{footnote}
The transition from on-surface adsorption to subsurface absorption is 
identified in experiment by a work function change, Refs.~\protect\onlinecite{barnard}
and \protect\onlinecite{palmer}.
\bibitem{footnote3}The choice of exchange flavor in vdW-DF
was set in Ref.~\protect\onlinecite{ijqc} to avoid artificial bonding in
noble-gas systems and to better mimic exact exchange calculations
for those systems. However, it is far from certain and even unlikely
that the conclusions drawn for noble-gas systems carry over to bonding 
separations smaller than 3 {\AA}.
\end{thebibliography}
\end{document}